\input harvmac
\input epsf.sty


\noblackbox

\font\cmss=cmss10
\font\cmsss=cmss10 at 7pt

\def\inbar{\vrule height1.5ex width.4pt depth0pt}

\def\IN{\relax{\rm I\kern-.18em N}}
\def\IB{\relax\hbox{$\inbar\kern-.3em{\rm B}$}}
\def\IC{\relax\hbox{$\inbar\kern-.3em{\rm C}$}}
\def\IQ{\relax\hbox{$\inbar\kern-.3em{\rm Q}$}}
\def\ID{\relax\hbox{$\inbar\kern-.3em{\rm D}$}}
\def\IE{\relax\hbox{$\inbar\kern-.3em{\rm E}$}}
\def\IF{\relax\hbox{$\inbar\kern-.3em{\rm F}$}}
\def\IG{\relax\hbox{$\inbar\kern-.3em{\rm G}$}}
\def\IGa{\relax\hbox{${\rm I}\kern-.18em\Gamma$}}
\def\IH{\relax{\rm I\kern-.18em H}}
\def\IK{\relax{\rm I\kern-.18em K}}
\def\IL{\relax{\rm I\kern-.18em L}}
\def\IP{\relax{\rm I\kern-.18em P}}
\def\IR{\relax{\rm I\kern-.18em R}}
\def\Z{\relax\ifmmode\mathchoice{\hbox{\cmss Z\kern-.4em Z}}{\hbox{\cmss Z\kern-.4em Z}} {\lower.9pt\hbox{\cmsss Z\kern-.4em Z}}{\lower1.2pt\hbox{\cmsss Z\kern-.4em Z}}\else{\cmss Z\kern-.4em Z}\fi}

\def\II{\relax{\rm I\kern-.18em I}}
\def\one{\relax{\rm 1\kern-.25em I}}

\def\S{Sec.~}

\def\CLL{\relax{\CL\kern-.74em \CL}}

\lref\Coleman{
  S.~R.~Coleman,
  ``Quantum sine-Gordon equation as the massive Thirring model,''
  Phys.\ Rev.\  D {\bf 11}, 2088 (1975).
}

\lref\disorder{
  T. Giamarchi and H. J. Schulz,
  ``Anderson localization and interactions in one-dimensional metals,''
 Phys. Rev. B {\bf 37}, 325 (1988).
}

\lref\qhz{
Xiao-Liang Qi, Taylor L. Hughes, and Shou-Cheng Zhang,
``Topological field theory of time-reversal invariant insulators,"
Phys. Rev. B {\bf 78}, 195424 (2008)
}

\lref\rsfl{
Shinsei Ryu, Andreas Schnyder, Akira Furusaki, Andreas Ludwig,
``Topological insulators and superconductors: ten-fold way and dimensional hierarchy,"
New J. Phys. 12, 065010 (2010).
}

\lref\kitaev{
Alexei Kitaev,
``Periodic table for topological insulators and superconductors,"
arXiv:0901.2686 [cond-mat].
}

\lref\anomaly{
Stephen L. Adler,
``"Axial-Vector Vertex in Spinor Electrodynamics,"
Phys. Rev. {\bf 177}, 2426 (1969);
J. S. Bell and R. Jackiw,
``"A PCAC puzzle: $\eta^0 \rightarrow \gamma \gamma$ in the $\sigma$-model"
 Nuovo Cim.A60:47-61,1969.
}

\lref\recent{
Carlos Hoyos-Badajoz, Kristan Jensen, Andreas Karch,
``A Holographic Fractional Topological Insulator,"
arXiv:1007.3253 [hep-th];
Andreas Karch, Joseph Maciejko, Tadashi Takayanagi,
``Holographic fractional topological insulators in 2+1 and 1+1 dimensions,"
arXiv:1009.2991 [hep-th]}

\lref\inter{
Lukasz Fidkowski and Alexei Kitaev,
``Effects of interactions on the topological classification of free fermion systems,"
Phys. Rev. B {\bf  81}, 134509 (2010);
Lukasz Fidkowski and Alexei Kitaev,
``Topological phases of fermions in one dimension,"
arXiv:1008.4138 [cond-mat].
}

\lref\thouless{
D. J. Thouless,
``Quantization of particle transport,"
Phys. Rev. B {\bf 27}, 6083 (1983).
}

\lref\goldstonewilczek{
Jeffrey Goldstone and Frank Wilczek,
``Fractional Quantum Numbers on Solitons,"
Phys. Rev. Lett. {\bf 47}, 986 (1981).
}

\lref\bla{
Erez Berg, Michael Levin, and Ehud Altman,
``Quantized pumping and phase diagram topology of interacting bosons,"
arXiv:1008.1590 [cond-mat].
}

\lref\apelrice{
W. Apel and T. M. Rice,
``Combined effect of disorder and interaction on the conductance of a one-dimensional fermion system,"
Phys. Rev. B {\bf 26}, 7063-7065 (1982).
}

\lref\stonemaslov{
Dmitri L. Maslov and Michael Stone,
``Landauer conductance of Luttinger liquids with leads,"
Phys. Rev. B {\bf 52} 5539 (1995).
}

\lref\thomaleseidel{
Ronny Thomale and Alexander Seidel,
``A minimal model of quantized conductance in interacting ballistic quantum wires,"
arXiv:1005.4228 [cond-mat].
}

\lref\umklapp{
F. D. M. Haldane,
``General Relation of Correlation Exponents and Spectral Properties of One-Dimensional Fermi Systems: Application to the Anisotropic S=1/2 Heisenberg Chain,"
Phys. Rev. Lett. {\bf 45}, 1358 (1980).
} 

\lref\blackemery{
J. L. Black and V. J. Emery,
``Critical properties of two-dimensional models,"
Phys. Rev. B {\bf 23}, 429 (1981).
} 

\lref\dennijs{
M. P. M. den Nijs,
``Derivation of extended scaling relations between critical exponents in two-dimensional models from the one-dimensional Luttinger model,"
 Phys. Rev. B {\bf 23}, 6111 (1981).
 } 

\lref\reviews{
E. Fradkin,
{\it Field Theories of Condensed Matter Systems}"
(Addison-Wesley, Redwood City, 1991);

I. Affleck, 
"Field Theory Methods and Quantum Critical Phenomena", Fields, Strings and Critical
Phenomena, p. 563-640, (ed. E. Brezin and J. Zinn-Justin North-Holland, Amsterdam, 1990), proceedings of
Les Houches Summer School, 1988, invited lectures.
}

\lref\aniso{
F. D. M. Haldane,
``Spontaneous dimerization in the S=1/2 Heisenberg antiferromagnetic chain with competing interactions,"
Phys. Rev. B {\bf 25}, 4925 (1982).
} 

\lref\grossneveu{
David J. Gross and Andre Neveu,
``Dynamical symmetry breaking in asymptotically free field theories,"
Phys. Rev. D {\bf 10}, 3235 (1974).
}

\lref\gravanom{
Shinsei Ryu, Joel E. Moore, Andreas W. W. Ludwig,
``Electromagnetic and gravitational responses and anomalies in topological insulators and superconductors,"
arXiv:1010.0936 [cond-mat].
}

\lref\amit{
Daniel J. Amit, Yadin Y. Goldschmidt, and G. Grinstein,
``Renormalisation group analysis of the phase transition in the 2D Coulomb gas, Sine-Gordon
theory and XY-model,"
J. Phys. A: Math. Gen. {\bf 13} 585 (1980).
}

\lref\fractional{
Joseph Maciejko, Xiao-Liang Qi, Andreas Karch, Shou-Cheng Zhang,
``Fractional topological insulators in three dimensions,"
arXiv:1004.3628 [cond-mat];
Brian Swingle, Maissam Barkeshli, John McGreevy, T. Senthil
``Correlated Topological Insulators and the Fractional Magnetoelectric Effect,"
arXiv:1005.1076 [cond-mat].
}

\lref\levinstern{
Michael Levin, Ady Stern,
``Fractional topological insulators,"
Phys. Rev. Lett. {\bf 103}, 196803 (2009).
}

\lref\higherbose{
F. D. M. Haldane,
``Luttinger's Theorem and Bosonization of the Fermi Surface,"
Proceedings of the International School of Physics "Enrico Fermi", Course CXXI "Perspectives in Many-Particle Physics" eds. R. A. Broglia and J. R. Schrieffer (North-Holland, Amsterdam 1994) pp 5-29,
arXiv:0505529 [cond-mat].
}

\lref\morestone{
Anupam Garg, V.P. Nair, and Michael Stone,
``Nonabelian Bosonization And Topological Aspects Of Bcs Systems,"
Annals Phys.173:149,1987;
Michael Stone, Frank Gaitan,
``Topological Charge And Chiral Anomalies In Fermi Superfluids,"
Annals Phys.178:89,1987. 
}

\lref\fujikawa{
Kazuo Fujikawa,
``Path Integral for Gauge Theories with Fermions,"
Phys. Rev. D {\bf 21}, 2848,(1980).
}

\lref\wen{
Xie Chen, Zheng-Cheng Gu, Xiao-Gang Wen,
``Classification of Gapped Symmetric Phases in 1D Spin Systems,"
arXiv:1008.3745 [cond-mat];
Zheng-Cheng Gu, Zhenghan Wang, Xiao-Gang Wen,
``A classification of 2D fermionic and bosonic topological orders,"
arXiv:1010.1517 [cond-mat].
}

\lref\turner{
Ari M. Turner, Frank Pollmann, Erez Berg,
``Topological Phases of One-Dimensional Fermions: An Entanglement Point of View,"
arXiv:1008.4346 [cond-mat].
}

\lref\rebbijackiw{
R. Jackiw and C. Rebbi,
``Solitons with fermion number $1/2$,"
Phys. Rev. D {\bf 13}, 3398 (1976).
}

\lref\poly{
A. J. Heeger, S. Kivelson, J. R. Schrieffer, W. -P. Su,
``Solitons in conducting polymers,"
Rev. Mod. Phys. {\bf 60}, 781Ð850 (1988).
}

\lref\franzquant{
M.M. Vazifeh, M. Franz,
``Quantization and $2\pi$ Periodicity of the Axion Action in Topological Insulators,"
arXiv:1006.3355 [cond-mat].
}

\lref\topanderson{
H.-M. Guo, G. Rosenberg, G. Refael, M. Franz,
``Topological Anderson Insulator in Three Dimensions,"
arXiv:1006.2777 [cond-mat].
}

\lref\luttinger{
J. M. Luttinger,
``An Exactly Soluble Model of a Many?Fermion System,"
J. Math. Phys. 4, 1154 (1963).
}

\lref\tomonaga{
Sin-itiro Tomonaga,
``Remarks on Bloch's Method of Sound Waves applied to Many-Fermion Problems,"
Prog. Theor. Phys. Vol. 5 No. 4 (1950).
}

\lref\fukane{
Liang Fu, C.L. Kane,
``Time Reversal Polarization and a $Z_2$ Adiabatic Spin Pump,"
Phys. Rev. B {\bf 74}, 195312 (2006).
}

\Title
{\vbox{
\baselineskip12pt
\hbox{MIT-CTP/4189}
}}
{\vbox{
\baselineskip22pt 
{\centerline{Interactions and the Theta Term}
\centerline{in One-Dimensional Gapped Systems} }}}

\centerline{Michael Mulligan}
\bigskip
{\it \centerline{Center for Theoretical Physics, MIT, Cambridge, MA 02139, USA}}

\medskip
\centerline{Email: mcmullig@mit.edu}

\bigskip
\noindent
We study how the $\theta$-term  is affected by interactions in certain one-dimensional gapped systems that preserve charge-conjugation, parity, and time-reversal invariance.  
We exploit the relation between the chiral anomaly of a fermionic system and the classical shift symmetry of its bosonized dual.
The vacuum expectation value of the dual boson is identified with the value of the $\theta$-term for the corresponding fermionic system.  
Two (related) examples illustrate the identification.  
We first consider the massive Luttinger liquid and find the $\theta$-term to be insensitive to the strength of the interaction.  Next, we study the continuum limit of the Heisenberg XXZ spin-1/2 chain, perturbed by a second nearest-neighbor spin interaction.  For a certain range of the XXZ anisotropy, we find that we can tune between two distinct sets of topological phases by varying the second nearest-neighbor coupling.  In the first, we find the standard vacua at $\theta = 0,\pi$, while the second contains vacua that spontaneously break charge-conjugation and parity with fractional $\theta/\pi = 1/2, 3/2$.  We also study quantized pumping in both examples following recent work.

\Date{2011}

\newsec{Introduction}

\subsec{Generalities}

Certain insulators can be distinguished by the presence or absence of a topological term in their low-energy effective actions \qhz.  In even spacetime dimensions the topological term is a $\theta$-term, while in odd dimensions, it is the Chern-Simons form.  (By topological, we mean a term that is independent of the background metric.)  More generally, non-interacting insulators are classified via the homotopy class of their gapped Hamiltonian  \refs{\kitaev,\rsfl}.

In this note, we are concerned with gapped systems defined in two spacetime dimensions and we study how the $\theta$-term depends upon certain types of interactions.  
In the continuum, the canonical example of a theory whose low energy action contains a $\theta$-term is provided by a single massive Dirac fermion $\psi$ minimally coupled to a background $U(1)$ gauge field $A_\mu$.  The Lagrangian has the form
\eqn\dirac{L = \int d^{2n}x \Big[\bar{\psi} (i\partial_\mu \gamma^\mu - e A_\mu \gamma^\mu - m)\psi \Big],
}
where $e,m$ are the charge and mass of the fermion, and $2n$ is the spacetime dimension.  The matrices $\gamma_\mu$ satisfy the algebra, $\{\gamma_\mu, \gamma_\nu \} = 2 \eta_{\mu \nu}$, with $\eta_{\mu \nu} = {\rm diag}(1, -1, ..., -1)$, and $\bar{\psi} := \psi^{\dagger} \gamma^0$.  Call the field strength of the gauge field, $F$.  Then, the $\theta$-term, obtained in the low energy Lagrangian for the above example, is given by the quantity
\eqn\gentheta{\theta I(F) =  {\theta \over 2} \int d^{2n} x (e F/2\pi)^n,
}
where $F^n := \epsilon^{i_1 j_1 ... i_n j_n} F_{i_1 j_1} \cdots F_{i_n j_n}$.
$\theta$ takes a value equal to either zero or $\pi$ depending upon the sign of the mass $m$.    When $\theta = 0$ the insulator is said to be trivial, while the insulator is called non-trivial when $\theta = \pi$.  

In general, $I(F)$ is integral when the theory is studied on a compact, oriented even-dimensional manifold.\foot{The factor of $1/2$ ensures that the minimum value of $I(F) = 1$ since we only consider systems on manifolds that admit a spin structure, i.e., fermions; otherwise, the $1/2$ should be removed.}  When the manifold is non-compact, quantization is a bit more subtle.  In two dimensions, this condition is equivalent to the assumption that the background $U(1)$ gauge field is compact and the energy of a gauge field configuration is finite.
In four dimensions, quantization of $I(F)$ follows from the assumption that the electromagnetic field configuration is produced by electrically and magnetically charged point particles -- Dirac quantization.\foot{For a recent discussion of the quantization of $I(F)$ in 3+1 dimension, please see \franzquant.}  
Quantization of $I(F)$ implies that $\theta$ has period $2\pi$ since it is the quantity $\exp(i \theta I(F))$ that appears in the path integral.  

For a given spacetime dimension, $ I(F)$ is odd under two of the discrete spacetime symmetries: charge-conjugation, parity, or time-reversal (${\cal C, P, T}$).  Given that $\theta$ has period $2 \pi$ and $I(F)$ is integral, the discrete symmetry under which $I(F)$ is odd folds the $2\pi$ interval in half; the two fixed points of the discrete symmetry are at $\theta =0, \pi$.  Thus, the positive and negative mass Dirac theories represent the two distinct classes of insulators invariant under the symmetries that change the sign of $I(F)$.  

To determine the discrete symmetries under which $I(F)$ is odd, it is necessary to know the transformation of the fields and coordinates.  Charge conjugation does not act upon the coordinates, however, it changes the sign of all components of the gauge field, $A_\mu \rightarrow - A_\mu$.  Parity is a reflection about the origin of all spatial directions $x_i \rightarrow - x_i$ and similarly reverses the sign of all spatial components of the gauge field $A_i \rightarrow - A_i$.  Time-reversal is an anti-unitary transformation, $i \rightarrow - i$, that transforms $t \rightarrow - t$ and $A_i \rightarrow - A_i$.  For example, in two and four spacetime dimensions, $I(F)$ transforms as $(-, -, +)$ and $(+,-,-)$, respectively, under $({\cal C, P, T})$, where $\pm$  refers to a quantity that is even/odd under the particular discrete symmetry.

In perturbation theory, a non-zero $\theta$-term is found in the effective action for a background gauge field coupled to a slowly varying charge density field after integrating out a massive Dirac fermion \goldstonewilczek.  The necessity of the charge density field should be contrasted with the situation in an odd number of spacetime dimensions.  In this case, no charge density field is required and the Chern-Simons form is found at one-loop by integrating out a massive Dirac fermion.  In odd dimensions, a component of the gauge field plays a role similar to that of the charge density field; this observation was used in the dimensional reduction technique of \qhz\ in deriving the existence of a $\theta$-term.  A second approach for determining the $\theta$-term for massive fermionic systems is reviewed in the paragraph below.  In \S2, we show more explicitly how a $\theta$-term is generated in a related example. 

The following argument relies upon the $U(1)$ chiral Adler-Bell-Jackiw anomaly \anomaly.
This argument is well known \refs{\qhz, \recent, \gravanom} and we review it for completeness.
Denote the product of all gamma matrices, $\gamma_{2n+1} : = i^n \gamma_0 ... \gamma_{2n-1}$, and complexify the fermion mass term as, 
\eqn\mass{m\bar{\psi} \psi = {1 \over 2} m \bar{\psi}(1 + \gamma_{2n+1})\psi + {1 \over 2} m^* \bar{\psi} (1 - \gamma_{2n+1} ) \psi.
}
Under a chiral transformation, $\psi \rightarrow \exp(i \alpha \gamma_{2n+1}) \psi$, the complex mass parameter rotates $m \rightarrow \exp(2 i \alpha) m$.  Because the path integration measure is not invariant under the chiral rotation, 
the quantum action also shifts by a term proportional to $2 \alpha I(F)$ \fujikawa.  (That the action for the Dirac fermion shifts by a $\theta$-term with $\theta = 2 \alpha$ is exact at one-loop.)  However, physical quantities cannot depend upon the choice of variables in the path integral.  Observables can only depend upon the product $M = m \exp(- i \theta)$ with $\theta$ being the coefficient of an $I(F)$ that may be present in the tree-level action.  Thus, the theory \dirac\ with positive mass is different from the one with negative mass with vanishing tree-level $\theta$ in both theories because $M$ differs by a sign.  However, it is possible to rewrite the Lagrangian for the negative mass theory by noting $-m = m \exp(i \pi)$, and so the two massive Dirac Lagrangians only differ by a term equal to $\pi I(F)$.  Once the massive fermion is integrated out, the effective Lagrangians for the background gauge field will differ by the presence of the term $\pi I(F)$.  

Please see \gravanom\ for a very nice discussion about how anomalies in various dimensions can be used to classify gapped systems.

\subsec{Motivation and Outline}

It is interesting to generalize these ideas to interacting topological phases \refs{\inter, \wen, \turner}.  
In the context of the massive Dirac theory, a generic interaction does not transform simply under a chiral transformation and so the above argument does not apply in general.   
In particular, a chiral rotation can change the sign of an interaction term and the low energy properties of a theory can be sensitive to the sign change. 
However, even if the interaction is invariant under a chiral transformation, the change of measure of the path integral, as inferred from an anomalous axial current, can naively depend upon the interaction.  
As an example of the latter phenomenon, in \S2 we study the 1+1 dimensional massive Luttinger liquid; here, the axial anomaly equation explicitly depends upon the interaction strength and the Fujikawa analysis is complicated by the presence of the marginal current-current interaction. 
In general, however, a relevant interaction introduces a new mass scale and may dramatically affect the low energy physics.

There are two approaches towards the study of interacting topological insulators.  The first is to consider a massive theory and add arbitrary interactions that maintain the symmetries of interest.  We do this in \S 2.   
The second approach is less direct and generally rather difficult to study in practice.  Begin with a classically gapless Lagrangian that admits a massive perturbation and ask whether or not the particular deformation results in a physically interesting topological phase.  We take this approach is \S3.  There need not be complete overlap with the various phases obtained (theoretically) via the above two approaches.  

Because the models we discuss and the techniques by which we study them are well known, let us summarize what is new.  Using bosonization to study interacting 1+1 dimensional fermionic systems, we identify the vacuum expectation value of the dual boson with the value of the $\theta$-term.  Bosonization allows us to determine the $\theta$-term directly.  We familiarize ourselves with this identification in the context of the massive Luttinger model.  Next, we relate the dimer and Neel phases of the Heisenberg XXZ spin-$1/2$ chain perturbed by a second nearest-neighbor interaction to the value of a corresponding $\theta$-term: the dimer and Neel phases contain the $\theta = \pi/2, 3\pi/2$ and $\theta = 0, \pi$ vacua, respectively.  Finally, we study adiabatic charge pumping \refs{\thouless, \fukane, \bla} in both models.  The novelty in the XXZ chain lies in the fact that it is necessary for the varied parameter to wind twice about the origin in order to transfer a full unit of charge across the system; a single winding is analogous to varying between a positive and negative mass theory in the context of the Luttinger liquid. 

This paper is organized as follows.  In \S2, we study the massive Luttinger liquid as an example of a 1+1 dimensional interacting topological phase.  We find that the $Z_2$ classification by the $\theta$-term is independent of the strength of the interaction.  In \S3, we discuss the continuum limit of the Heisenberg XXZ spin-$1/2$ chain, perturbed by a second nearest-neighbor interaction.  For a certain range of the XXZ anisotropy, we find that we can tune between two distinct sets of topological phases by varying the second nearest-neighbor coupling.  In the first set, we find the standard $\theta = 0,\pi$ vacua, while the second set contains vacua that spontaneously break charge-conjugation and parity with fractional $\theta = \pi/2, 3\pi/2$.  Adiabatic variation of certain parameters in both models results in charge transport.  In contrast to the massive Luttinger liquid \refs{\thouless, \bla}, the XXZ chain transitions to the $\theta = \pi$ vacuum after a single closed orbit in parameter space and only results in an half-unit of charge transfer.  Two windings are necessary in order to return to the original vacuum and so the total charge transferred is integral.
We summarize in \S 4.

\newsec{Massive Luttinger Liquid}

\subsec{Invariance of $\theta$}

A simple example where the interactions can be treated exactly is provided by the massive Luttinger liquid \refs{\tomonaga,\luttinger,\Coleman}.    
The Lagrangian
\eqn\masslutt{L = \int d^2x \Big[ \bar{\psi} (i\partial_\mu \gamma^\mu - e A_\mu \gamma^\mu)\psi - {1 \over 2} \bar{\psi}\Big(m(1+\gamma_3) + m^*(1-\gamma_3)\Big)\psi - {g \over 2} (\bar{\psi} \gamma_\mu \psi)(\bar{\psi} \gamma^\mu \psi) \Big].
}
\masslutt\ describes a 1+1 dimensional spinless fermion $\psi$ of charge $e$ coupled to a background gauge field $A_\mu$.  The constant charge density field $m$ couples the left and right handed components of $\psi$ and functions as the mass parameter.  The Lagrangian breaks both ${\cal C}$ and ${\cal P}$ when $m$ has a non-zero imaginary part.  The current-current interaction is parameterized by the coupling $g$.  A more general quartic interaction will be considered in \S3.  

At finite $g$, it is simplest to study \masslutt\ after bosonization of the fermion \Coleman; we introduce a scalar field $\phi$ that satisfies
\eqn\boserules{\eqalign{1+ g/\pi & = 4\pi/\beta^2, \cr
\bar{\psi} i \partial_\mu \gamma^\mu \psi & = {1 \over 2} (\partial_\mu \phi)^2, \cr
\bar{\psi} \gamma^\mu \psi & = {\beta \over 2\pi} \epsilon^{\mu \nu} \partial_\nu \phi, \cr
{1 \over 2} \bar{\psi}(1+\gamma_3)\psi & =  \psi^{\dagger}_R \psi_L =  {1 \over 2 \beta^2} \exp(i \beta \phi),
}}  
where the magnitude of the coefficient of the exponential in the final equality in \boserules\ is chosen for convenience.  Substituting \boserules\ into \masslutt, and after one integration by parts in the last term, we obtain
\eqn\massbose{L = \int d^2x \Big[ {1 \over 2} (\partial_\mu \phi)^2 - {m \over 2 \beta^2} \exp(i\beta \phi) - {m^* \over 2 \beta^2} \exp(-i \beta \phi) + {\beta e \over 2 \pi} \phi \epsilon^{\mu \nu} \partial_\mu A_\nu\Big].
}
The integration by parts in the last term of \massbose\ is required if the action is to be gauge-invariant on a space with boundary.  Further, this form of the action correctly displays the chiral anomaly associated with shifts of the $\phi$ field or, in the fermionic language, the Jacobian of the path integration measure under a chiral transformation.  More precisely, a chiral rotation of $\psi \rightarrow \exp(i \alpha \gamma_3) \psi$ is equivalent to a shift of $\phi$ by $\alpha$ and the last term in \masslutt\ correctly reproduces the shift of the action by $I(F)$ under this transformation.  We also note that $\phi \rightarrow - \phi + \pi$ under a ${\cal C}$ transformation, while $\phi$ changes sign under a ${\cal P}$ transformation, and is neutral under ${\cal T}$.  (Although ${\cal C}$ effectively takes $m \rightarrow - m$, it also shifts $\theta$, thus preserving $m \exp(- i\theta)$.)

Expanding the exponential $\exp(\pm i \beta \phi)$, $\beta^2$ acts as a coupling constant for the leading quartic scalar interaction, assuming for the moment that $m$ is real.  Thus, the duality \boserules\ interchanges a strongly coupled fermion theory for a weakly coupled boson theory and vice versa.  When $\beta^2 = 4 \pi$, $g$ vanishes and \massbose\ relates a free massive fermion to an interacting scalar field.  Through its definition in terms of $\beta$, the interaction strength $g$ of the four-fermion interaction controls the radius of the $\phi$ boson; $\phi$ is periodically identified under $\phi \rightarrow \phi + 2\pi/\beta$.  Invariance of $\exp(i L)$ under a shift of $\phi$ by its period is consistent with the integrality of $I(F)$; in general, however, integrality of $I(F)$ is distinct from the requirement that $\exp(i L)$ be invariant under periodic shifts of $\phi$.  For the remainder, we restrict $\phi$ to the interval $[0, 2\pi/\beta)$.       

The scaling dimension of $\exp(i \beta \phi)$ equals $\beta^2/4\pi$ or $(1+g/\pi)^{-1}$ in terms of the fermionic interaction strength $g$.  Attractive interactions  between the fermions increase the dimension of the exponential operator, while repulsive interactions have the opposite effect.  $\beta^2 \geq 8\pi$ or $g \leq - \pi/2$ parameterize a line of fixed points that terminates when $\beta^2 = 8\pi$ at the $SU(2)_1$ WZW critical point \amit.  Thus, we restrict to the regime where $\exp(i \beta \phi)$ is relevant.

It is convenient to absorb the phase of the mass $m$ by redefining $\phi$.  Writing $m = |m| \exp(i a)$, we rename $\tilde{\phi} = \phi + a/\beta$.  With this redefinition, the Lagrangian becomes
\eqn\massbose{L = \int d^2x \Big[ {1 \over 2} (\partial_\mu \tilde{\phi})^2 - {|m| \over \beta^2} \cos(\beta \tilde{\phi})+ {\beta e \over 2 \pi} \Big(\tilde{\phi} - a/\beta \Big) \epsilon^{\mu \nu} \partial_\mu A_\nu  \Big].
}
At energies low compared to $|m|$ the scalar field is frozen and minimizes its potential energy at $\langle \tilde{\phi} \rangle = \pi/\beta$ or, in the original coordinate, $\langle \phi \rangle = (\pi - a)/\beta$.  In the low energy effective Lagrangian
\eqn\lowl{L_{\rm eff} = {(\pi - a) \over 2} \int d^2x (e\ \epsilon^{\mu \nu} F_{\mu \nu}/2\pi),
} 
the dependence upon $\beta$ has completely disappeared; the effective Lagrangian is independent of the strength of the current-current interaction.  We see explicitly that the low energy Lagrangian equals $(\pi - a) I(F)$.  The ${\cal C}$ and ${\cal P}$ symmetric points lie at $a = 0, \pi$; namely, when $m>0$ or $m<0$.\foot{We remark that identifying the $\pm m$ theories with the $\theta = 0, \pi$ vacua is a convention. An invariant statement requires a specification of a tree-level $\theta$-term -- tuned to zero in this example -- as discussed in the Introduction. In the absence of this, we can only say that the effective actions for the two theories differ by a $\theta = \pi$ term.}  When $a$ differs from these two values, the theory breaks ${\cal C}$ and ${\cal P}$.  These results are independent of the interaction strength.

\subsec{Adiabatic Variations}

In this section, we re-derive the results recently obtained in \bla, and discussed previously in slightly more generality in \refs{\fukane, \qhz}, concerning the Thouless charge pump \thouless.  This review provides intuition for a similar study in the context of the Heisenberg XXZ spin-$1/2$ chain discussed in \S3.  As an aside, we mention how the chiral anomaly equation is related to the conductance of a one-dimensional wire.    

It is natural to consider two currents constructed from the fields in \masslutt: (1) the vector current $J_V^\mu = \bar{\psi} \gamma^\mu \psi$ and (2) the axial current $J_A^\mu = \bar{\psi} \gamma_3 \gamma^\mu \psi$.  Their integrated charge densities describe, respectively, the sum and difference of the number of left and right moving fermions.  If both currents were conserved, the number of left and right moving fermions would be separately conserved.  However, the axial current is not conserved because of the fermion mass term and the minimal coupling to the background gauge field.   The bosonization rules \boserules\ allow us to rewrite these currents in terms of the scalar $\phi$: $J_V^\mu = {\beta \over 2 \pi} \epsilon^{\mu \nu} \partial_\nu \phi$ while $J_A^\mu = {\beta \over 2 \pi} \partial^\mu \phi$.    The vector current is identically conserved while the axial current, given its definition in terms of $\phi$, obeys an equation that follows directly from the $\phi$ equation of motion,
\eqn\axial{\partial_\mu J_A^\mu = {|m| \over 2 \pi} \sin(\beta \tilde{\phi}) + {\beta^2 e \over (2 \pi)^2} \epsilon^{\mu \nu} \partial_\mu A_\nu.
}
In the fermion language, the first term on the righthand side of \axial\ is the explicit violation of the axial symmetry by the mass term while the second term obtains from a variation of the fermion path integration measure.  Note that the coefficient of $\epsilon_{\mu \nu} \partial_\mu A_\nu$ depends upon the interaction through $\beta^2/4\pi$.

Consider first an adiabatic change of sign of the fermion mass when the background field is set to zero.  In particular, consider the arc parameterized at non-zero $|m|$ by adiabatically varying the phase $a$ with respect to time.  In the vacuum, the vector current $\langle J_V^\mu \rangle = {\beta \over 2\pi} \epsilon^{\mu \nu} \langle \partial_\nu \phi \rangle = - {1 \over 2\pi} \epsilon^{\mu \nu} \partial_\nu a$.  Therefore, the amount of charge that passes through an arbitrary point on the line,
\eqn\charge{\Delta Q = e \int_{- \infty}^{\infty}dt J_V^x = {e \over 2\pi} \int_{- \infty}^{\infty} dt\ \partial_t a = {e \over 2\pi} \Delta a.
}
Choosing $\Delta a = 2 \pi$, i.e., a loop about the origin, \charge\ is simply a restatement of Thouless' result on quantized charge transport in one spatial dimension \thouless\ generalized to the Luttinger liquid as nicely demonstrated recently in \bla. 
Adiabatic variation of the phase of the mass term about the origin results in transport of a single unit of charge (modulo one) through the system.  If the loop does not enclose the origin, $\Delta a = 0$ and no charge is transported.  \charge\ gives an explicit example with interactions of the phenomenon discussed in \qhz\ where two distinct topological insulators are connected via a path in parameter space that preserves a non-zero excitation gap, but does not necessarily maintain the discrete symmetries of the two classes of insulator.  

Contact with the work of \refs{\rebbijackiw, \poly, \goldstonewilczek} can be made by considering instead an interpolation in space instead of time between two values of $m$.  It is sufficient to consider an interpolation of the phase $a$ of $m$.  The charge carried by a soliton,
\eqn\chargesoliton{Q =   e \int dx J^0_V = e \int dx {\beta \over 2\pi} \partial_x \langle \phi \rangle = {e \over 2\pi} \Delta a.
} 
If the soliton interpolates between positive and negative values of the mass parameter, then $\Delta a = \pi$ and so $Q = e/2$  \refs{\rebbijackiw}.  However, in general, the change of the phase of $m$, and therefore the corresponding charge carried by the soliton, can be arbitrary \goldstonewilczek.  

This provides an heuristic way to understand why the charge transported for any closed loop in $m$ space is integral.  Consider a finite length chain and imagine introducing a soliton or kink near the left end that interpolates between the two vacua at $\pm m$.  This interpolation is achieved by varying the phase of $m$ as above in space as opposed to time.  The charge carried by the soliton, $Q = e/2$. Closing the loop by returning to the positive $m-$axis results in the transport of a second $e/2$ soliton.  If we had instead reversed the path in $m$ space so that the origin was not enclosed, it would have been equivalent to taking a kink from the left to the right end and then back again with zero net charge transfer.

As an aside, we mention a result that follows from the axial current equation \axial.  Set $m=0$ and allow the external electric field $\epsilon^{\mu \nu} \partial_\mu A_\nu$ to vary adiabatically with time.  In particular, adiabatically vary $A_x$ while keeping $A_t = 0$.  The system is placed on a spatial circle of length $L$ and $\psi$ is taken to be periodic around $L$ (more general boundary conditions are also possible).
A general constant  background $A_x$ cannot be removed by a gauge transformation as this would ruin the assumed periodicity of $\psi$ since a gauge transformation multiplies the fermion field by $\exp(i e \alpha)$ with $\alpha$ the gauge transformation parameter.
However, when $A_x = 2\pi n/e L$, for $n$ an integer, it is possible to perform a gauge transformation with the choice $\alpha = A_x x$ to remove the background field.  In other words, a gauge field configuration $A_x = 2\pi n / e L$ is physically equivalent to a configuration without any background field.  

Let $A_x$ adiabatically vary from zero to $2\pi/eL$ so that the system returns to its original ground state after the variation with, perhaps, a slight relabeling of its spectrum.
Integrating \axial\ with $m=0$ over spacetime, we find
\eqn\index{N_L - N_R = \int dt dx\ \partial_\mu J_A^\mu = {\beta^2 e \over 4 \pi^2} \int dt dx\ \partial_t A_x = {\beta^2 e L \over 4 \pi^2} \Delta A_x.
}
Because $\Delta A_x = 2\pi/e L$, the difference $N_L - N_R$ changes by $\beta^2 \over 2\pi$.  Note that if the interaction $g$ is tuned to zero, $\beta^2 = 4 \pi$ and the difference $N_L - N_R = 2$ reflecting the fact that one left-handed fermion has emerged from the fermi sea, while one right-handed fermion has been captured.  Such a flow in the spectrum ensures the charge of the vacuum remains the same after the adiabatic variation.  However, a non-zero current obtains after the variation since there is an imbalance between the number of left and right movers.  For arbitrary interaction, that the difference is proportional to $\beta^2/4\pi$ is a manifestation that the conductance of an interacting 1+1 dimensional chain equals ${\beta^2 \over 4\pi} {e^2 \over h}$ (assuming Luttinger liquid leads) \refs{\apelrice, \stonemaslov, \thomaleseidel}.

\newsec{Generic Interactions}

In this section, we study slightly more generic interactions that preserve ${\cal C}$ and ${\cal P}$.  We focus on a one-dimensional system that is gapped in the IR, not because of a tree-level mass term, but because of interactions.\foot{See \topanderson\ for a related discussion in 3+1 dimensions where the interactions are due to disorder.  In general, disorder may be treated exactly in 1+1 dimensions \disorder.}  We view this as an alternative approach to building insulators with interesting topological properties.

\subsec{$\theta$-Vacua}

The motivation for the model considered in this section is the Heisenberg XXZ spin-$1/2$ chain described by the microscopic Hamiltonian,
\eqn\spinchain{H = J \sum_{i} \Big[ S^x_i \cdot S^x_{i+1} + S^y_i \cdot S^y_{i+1} + g' S_i^{z} \cdot S_{i+1}^z\Big],
}
with $J, g'>0$.  The spin operators obey the $SU(2)$ algebra, $[S^a_j, S^b_j] = i \epsilon^{abc} S^c_j$ for spins on the same lattice site, $j$.  $g'$ parameterizes a line of critical points for $0 \leq g' \leq 1$.  The system enters a massive phase when $g'>0$.  (The system is ferromagnetic when $g'<0$.)  It is well known that the continuum limit is described by the Lagrangian (for a review, see \reviews),
\eqn\continuum{L = \int d^2x \Big[ \bar{\psi} (i\partial_\mu \gamma^\mu - e A_\mu \gamma^\mu)\psi - {g \over 2} \Big((\bar{\psi} \gamma^\mu \psi)^2 - 2((\psi_R^{\dagger} \psi_L)^2 + (\psi_L^{\dagger} \psi_R)^2) \Big],
}
where the subscripts $L,R$ refer to the left and right moving parts of the fermion.  In obtaining \continuum, we have dropped unimportant constants and absorbed any renormalization of the Fermi velocity by a redefinition. Note that we have introduced in \continuum\ a background gauge field minimally coupled to the number current of the Jordan-Wigner fermions.  Fermion number is identified with the $S_z$ eigenvalue in the spin language.  The interaction terms are parameterized by the coupling $g$ which is linearly related to the XXZ anisotropy parameter, $g'$.  In \continuum, the first interaction term is simply the current-current interaction whose effects we previously studied in \S2.  The second interaction term, naively zero by the Pauli principle, is responsible for dynamical mass generation.  The product should be understood as the leading term in the operator product expansion (OPE) of the operator $(\psi^{\dagger}_{R,L} \psi_{L,R})$ with itself.  This second term is the spin-less analog of Umklapp scattering \refs{\umklapp, \blackemery, \dennijs} and we refer to it as the Umklapp term.

Bosonization proceeds as in \S2 by introducing a scalar field $\phi$ satisfying \boserules.  The resulting Lagrangian
\eqn\general{L = \int \Big[ {1\over 2} (\partial_\mu \phi)^2  + {g \over \beta^2} \cos(2 \beta \phi) + {\beta e \over 2 \pi} \phi \epsilon^{\mu \nu} \partial_\mu A_\nu\Big].
}
We assume that we are in the regime $\beta^2 < 2\pi$ where the Umklapp term is relevant.  (The precise value in terms of microscopic parameter $g'$ of the XXZ chain at which this transition occurs can be obtained and supports the qualitative prediction of the sine-Gordon model that a transition to a massive phase occurs for some finite $g \sim \pi$.)  
The low energy limit of \general\ is achieved by taking $g \rightarrow \infty$.  The Umklapp potential freezes the scalar field at one of the two vacua, $\langle \phi \rangle = 0, \pi/\beta$.   These two vacua are $Z_2$ chiral partners and correspond to $\theta = 0, \pi$, respectively.

Notice that in contrast to the example in \S2 where the tree level mass term only allowed a single vacuum for a particular sign of the cosine interaction (assuming the range of $\phi$ was restricted to the interval $[0, 2\pi/\beta)$), the two topologically distinct vacua for this model both occur for one sign of the cosine interaction.  Of course, this is due to the doubled periodicity of the cosine potential.  Note there do not exist additional non-trivial vacua when $g<0$.  The reason is that the dimension of the Umklapp operator is determined by $g$ and so when $g<\pi$, $\cos(2 \beta \phi)$ is irrelevant and decouples from the low energy physics.  

It is possible to find additional vacua when the coefficient of the cosine is negative, however, it is necessary to independently vary the dimension of the Umklapp operator and its coefficient.
In terms of the motivating XXZ chain, this is achieved by adding a second nearest-neighbor spin interaction \aniso,
\eqn\secondnn{\delta H = \lambda \sum_{i} {\bf S}_i \cdot {\bf S}_{i+2}.
}  
The bosonic version of the continuum Lagrangian
\eqn\generalnn{L = \int \Big[ {1\over 2} (\partial_\mu \phi)^2  + {(g - 6 \lambda) \over \beta^2} \cos(2 \beta \phi) + {\beta e \over 2 \pi} \phi \epsilon^{\mu \nu} \partial_\mu A_\nu\Big],
}
where $\beta^2/4\pi = (1 + (g + 2 \lambda)/\pi)^{-1}$.  Choose $g>\pi$ so that the Umklapp operator is relevant even when the second nearest-neighbor coupling vanishes.  
The massless Luttinger liquid separates at $\lambda = g/6$ two distinct sets of massive phases .
When $\lambda < g/6$, the vacua lie at $\langle \phi \rangle = 0, \pi/\beta$, while when $\lambda > g/6$, $\langle \phi \rangle = \pi/2\beta, 3\pi/2\beta$ and the corresponding $\theta = \pi/2, 3\pi/2$.  
The two vacua in the regime $\lambda>g/6$ spontaneously break ${\cal C}$ and ${\cal P}$.
This follows from the fact that $I(F)$ is integral.  

If $I(F)$ were even, for example, the range of $\theta$ would be reduced to the interval $[0, \pi)$ and the two vacua at $\theta = \pi, 3\pi/2$ would be identified with their $\pi/\beta$ partners and effectively removed.  The remaining vacua would become the two ${\cal C, P}$ invariant vacua -- the vacuum with non-zero vacuum expectation value having fractional $\theta = \pi/2$.  We are not aware of an argument that would require $I(F)$ to be even.  Note that invariance of the total action with respect to a shift of $\phi$ that preserves the cosine interaction is equivalent to the requirement that $I(F)$ be even, however, quantization of $I(F)$ and the shift symmetry of a potential need not be related.

In terms of the spin-chain, the transition at $\lambda = g/6$ separates two distinct sets of phases.  The vacua that preserve ${\cal C, P}$ correspond to the Neel phase while the dimerized phase occurs for $\lambda > g/6$, given a sufficiently large $g>\pi$.  

Finally, we remark that \generalnn\ contains the most relevant interactions consistent with the symmetries of the spin chain.  One possible operator of lower dimension is $\cos(\beta \phi)$, obtained from an alternating interaction in the spin model, but it can be ignored since it violates the discrete $Z_2$ lattice symmetry, $\phi \rightarrow \phi/\beta$.  There is one other possibly dangerous operator, namely, $\cos(\beta \tilde{\phi})$ where $\tilde{\phi}$ is defined as follows.  Expand $\phi = \phi_L + \phi_R$ into left and right moving modes.  We define $\tilde{\phi} := \phi_L - \phi_R$.  (It can be identified with the T-dual of $\phi$.)  It is possible to ignore $\cos(\beta \tilde{\phi})$ as long as the spin-chain Hamiltonian \spinchain\ enjoys the $U(1)$ rotation symmetry about the $z$-axis since this acts as $\tilde{\phi} \rightarrow \tilde{\phi} + {\alpha}$  where $\alpha$ is an arbitrary constant.

\subsec{Adiabatic Variation}

Following the method discussed in \S2, it is now possible to examine the nature of charge transport under adiabatic variation of parameters in the XXZ spin-chain.  Just as it was necessary to allow the mass parameter in \S2 to be complex, we take the coefficient of the cosine interaction discussed in this section to be complex.  This is done as follows. 

 The continuum limit of the XXZ spin-chain is parameterized by two couplings, $g, \lambda$ which we allow to be complex.  $g+2\lambda$ determines the dimension of the cosine interaction while and $g-6\lambda$ is the coupling constant of the interaction.  Thus, we must keep the linear combination $g+2\lambda$ real while allowing $g-6\lambda$ to take complex values.  Hermiticity of the Lagrangian is maintained by adding in the Hermitian conjugate interaction.  Writing $g = g_1 + i g_2$ and $\lambda = \lambda_1 + i \lambda_2$, for $g_i, \lambda_i$ real, we consider arbitrary values of $g_1 - 6 \lambda_1 + i(g_2 - 6 \lambda_2)$ while requiring $g_2 = - 2 \lambda_2$ so that the dimension of the operator is real.  The relation between $g_2, \lambda_2$ reduces the dimension of the parameter space by one.  We take $g_1, g_2, \lambda_1$ to parameterize the continuum limit of the XXZ chain.  
 
 In order to simplify the presentation, let us define, $\tilde{g} = g_1 + 2 \lambda_1$ and $\rho \exp(i a ) = g_1 - 6 \lambda_1 + 4 i g_2$, with $0 \leq \rho = \sqrt{(g_1 - 6 \lambda_1)^2 + 16 g_2^2}$ and $a = \tan^{-1}(4g_2/(g_1-6\lambda_1))$.  The resulting bosonic Lagrangian 
 \eqn\final{L = \int d^2 x \Big[ {1 \over 2} (\partial_\mu \phi)^2 + \rho \cos( 2 \tilde{\beta} \phi + a) + {\tilde{\beta} e \over 2 \pi} \phi \epsilon^{\mu \nu} \partial_\mu A_\nu \Big],
 }
where $4\pi/\tilde{\beta}^2 = 1 + \tilde{g}/\pi$.  At low energies or $\rho \rightarrow \infty$, the vacua lie at $\langle \phi \rangle = - a/2\tilde{\beta} + n\pi/\tilde{\beta}$, for integer $n$. 
 
We consider the path obtained at non-zero $\rho$ by varying $a: 0 \rightarrow 2\pi$.  Following \S2, the amount of charge that is transferred across the system 
\eqn\chargegeneral{\Delta Q = - e \int_{- \infty}^{\infty}dt {\tilde{\beta} \over 2\pi} \partial_t \langle \phi \rangle = {e \over 2\pi} \int_{- \infty}^{\infty} dt\ \partial_t a = {e \over 4\pi} \Delta a.
}
The factor of $1/2$ on the right hand side of \chargegeneral\ is due to the doubled periodicity of the cosine interaction.  When $\Delta a = 2\pi$, an half-integral charge is transferred across the system.  The system, however, has not returned to its initial state, but has transitioned to its $Z_2$ chiral partner.  It is necessary to wind twice about the origin, $a: 0 \rightarrow 4\pi$, in order to return to the original vacuum.  Doing so, we find integral charge transport.

\newsec{Discusion}

We considered the effects of certain interactions on the $\theta$-term for one-dimensional gapped systems.  Our examples were simple and well known. What is new is our viewpoint on the transitions described by tuning of various parameters in the models.  We made use of abelian bosonization techniques to determine the value of the $\theta$-term in the low energy action.  In particular, the semi-classical minimum of the potential for the dual boson was shown to directly determine the value of the $\theta$-term.

The first model we studied was that of the massive Luttinger liquid.  When the interactions are taken to vanish, the model describes free massive fermions.  The sign of the mass determines the topological phase in which the system lies. The two phases are distinguished by the value of a $\theta$-term.  We found these two values of $\theta$ to be stable to arbitrary current-current interactions.  

In the course of studying this model, we made contact with the recent work \bla.  By adiabatically varying the phase of the mass term, a unit of charge is pumped through the system.  In the massless limit, we also noticed that the axial anomaly equation provided a simple way to infer the conductance of a one-dimensional interacting wire.

The second model we studied was inspired by the Heisenberg XXZ spin-$1/2$ chain perturbed by a second nearest-neighbor interaction.  For vanishing second nearest-neighbor interaction, we found two vacua related to each other by a $Z_2$ chiral translation.  Phases with fractional $\theta/\pi = 1/2, 3/2$ obtained when this interaction was turned on with sufficient strength, however, these vacua spontaneously broke both ${\cal C}$ and ${\cal P}$.  In XXZ chain, we studied adiabatic charge pumping and found, in contrast to the model studied in \S2, it was necessary to perform two cycles in order to return to the same vacuum at the end of the variation.  Again, an integral unit of charge was pumped during the variation.  

Part of the initial motivation in studying simple one-dimensional systems amenable to bosonization techniques was to have a better understanding of the $\theta$-term in the context of various topological phases of matter.  In particular, we were interested in studying interactions in the controlled environs of one-dimensional systems.  
This work provides context for further exploration of phases of matter where we can hope to find fractional values of $\theta/\pi$ preserving ${\cal C, P, T}$ \refs{\fractional, \recent}.

\bigskip

\eject 
\centerline{\bf{Acknowledgements}}

It is a pleasure to thank F. Benini, F. Burnell, K.-T. Chen, C. Herzog, C. Hoyos-Badajoz, S. Kachru, P. Lee, J. McGreevy, A. Scardicchio, and B. Swingle for discussions and encouragement.
We  acknowledge the hospitality of the Stanford Institute for Theoretical Physics, the Galileo Galilei Institute for Theoretical Physics, and Oxford University while this work was in progress.  We were supported in part by funds provided by the U.S. Department of Energy (D.O.E.) under cooperative research agreement DE-FG0205ER41360.

\listrefs
\end